\documentclass[11pt,onecolumn,amssymb,nofootinbib]{revtex4}
\usepackage{amsmath, amsthm, amscd, amssymb}
\usepackage{bm}
\usepackage{bbm}

\begin{document}

\title{\bf Hubble parameter and related formulas for a Weyl scaling invariant dark energy action}

\author{Stephen L. Adler}
\email{adler@ias.edu} \affiliation{Institute for Advanced Study,
Einstein Drive, Princeton, NJ 08540, USA.}

\begin{abstract}
We extend our previous analysis of a model for ``dark energy'' based on a Weyl scaling invariant dark energy action.   We reexpress all prior results in terms of proper time, using the fluctuation amplitude $\Phi$ without approximation, and derive a compact formula for the squared effective Hubble parameter. This formula involves effective dark energy and matter densities that differ from their expressions in the standard $\Lambda CDM$ cosmology.  We also give new analytic results for the function $\Phi$  and discuss their implications.
\end{abstract}

\maketitle
\section{Introduction}
In a series of papers over the last eight years we have explored the postulate that the part of the gravitational action that depends only on the undifferentated metric $g_{\mu\nu}$, but involves no metric derivatives, is invariant under the Weyl scaling $g_{\mu\nu}\to \lambda g_{\mu\nu}$. Adoption of this postulate implies that the so-called  ``dark energy'' cannot be a vacuum energy with action
\begin{equation}\label{cosm}
S_{\rm cosm}=-\frac{\Lambda}{8 \pi G} \int d^4x ({}^{(4)}g)^{1/2}~~~,
\end{equation}
sicnce this is not Weyl scaling invariant and so must have coefficient zero. The simple modification of dividing the integrand by $g_{00}^2$ gives
 a Weyl scaling invariant, three-space general coordinate invariant,  but frame-dependent,  gravitational action
\begin{equation}\label{eff}
S_{\rm eff}=-\frac{\Lambda}{8 \pi G} \int d^4x ({}^{(4)}g)^{1/2}(g_{00})^{-2}~~~.
\end{equation}
Since the unperturbed Freedman-Lema\^itre-Robertson-Walker (FLRW) cosmological metric has  $g_{00}=1$, in this context the action of Eq. \eqref{eff} mimics the action of Eq. \eqref{cosm}, but when $g_{00}$ deviates from unity, their consequences differ. Our series of papers has been devoted to exploring the possibility that the ``dark energy'' that leads to accelerated expansion of the universe is really a consequence of the action of Eq. \eqref{eff}, rather than the action of Eq. \eqref{cosm}.  As a prelude to outlining what is new in this paper, we first give a mini-review of our earlier work.

\begin{itemize}
\item {\bf Motivations}  We originally formulated our postulate in \cite{adler1a}, motivated by Weyl scaling  properties of massless field theories and implications of this for the gravitational effective action. We were also motivated by the fact that the cosmic microwave background radiation specifies a preferred Lorentz frame, suggesting that frame dependence could play a role in fundamental physics at some level.    In a  Gravitational Essay \cite{adler1b} giving an exposition of our proposal, we noted that 't Hooft, also in a Gravitational Essay \cite{'thooft}, had proposed that Weyl scaling invariance is ``the missing symmetry component of space and time'', reinforcing  interest in this form of scaling as applied to the
    part of the gravitational action that is independent of derivatives. Later on, in a seminar talk at Princeton University \cite{prn}, we noted that if ``dark energy'' arises from the action of Eq. \eqref{eff}, then although there is a ``hierarchy problem'' to explain the very small value of $\Lambda$, there is not necessarily a ``fine-tuning problem''.  The cosmological constant fine-tuning problem arises when dark energy is a residual vacuum energy left over after almost exact cancellations of bosonic and fermionic vacuum energies.  An exact cancellation could be understood as the result of a selection rule, such as the Weyl scaling invariance that we have proposed, which requires the coefficient in the action of Eq. \eqref{cosm} to vanish.  An almost exact cancelation, to one part in $10^{120}$ in the absence of supersymmetry, or one part in $10^{60}$ if  there is supersymmetry broken at the TeV scale, corresponds to a fine tuning that is much harder to understand.  So it is important to see if the observed cosmological constant really arises as a vacuum energy from the action of Eq. \eqref{cosm}, or arises from the very different action of Eq. \eqref{eff},  which mimics a cosmological constant in unperturbed FLRW cosmology.

\item{\bf Equations of Motion, and Relation to Other Work with Broken Time Diffeomorphism}

    The total action in our model consists of taking Eq. \eqref{eff} as the cosmological constant part of the action, together with the Einstein-Hilbert action and the usual matter action.  Thus the different components of the action have conflicting invariance properties: the cosmological part is Weyl scaling invariant but only three-space general coordinate invariant, while the Einstein-Hilbert action and matter action are four-space general coordinate invariant, but not Weyl scaling invariant.  Since the cosmological part of the action is not four-space general coordinate invariant, the standard argument \cite{weinold} used to show that variation of an invariant action leads to a covariantly conserved energy momentum tensor does not apply.  So if we vary the full action with respect to all components of the metric $g_{\mu\nu}$, we  get inconsistent equations of motion:  the variation of the Einstein-Hilbert action gives the Einstein tensor $G_{\mu\nu}$, which obeys the Bianchi identity $D^{\mu}G_{\mu\nu}=0$, and the variation of the matter action gives the covariantly conserved matter energy-momentum tensor, while the variation of the cosmological action of Eq. \eqref{eff} gives a non-covariantly conserved tensor.  How can an action principle lead to inconsistent equations of motion?  The answer is that the ten components of $g_{\mu\nu}$ are not all independent variables; the Bianchi identities give four constraints, so only six components of the metric tensor are independent. Only when the total action is four-space generally covariant can one consistently vary the action with respect to all metric components, treating them as if they were independent.    So we have suggested \cite{adler1a} a procedure to get consistent equations of motion when Eq. \eqref{eff} is used for the dark energy component.  Following the ADM \cite{adm} formulation of the initial value problem in general relativity, we take the six spatial components $g_{ij}$ of the metric tensor as the independent variables, and vary the full action with respect to these only.  This gives as the Euler-Lagrange equation a tensor equation of the form $T^{ij}=0$.\footnote{Since the equations of motion for the independent variables $g_{ij}$ come from the extremum of a classical action,  we expect the usual heuristic arguments for stability of a quantized version, obtained from a Feynman path integral over this action,  to hold. } Then we integrate the covariant conservation condition $D_{\mu}T^{\mu\nu}=0$ to give the components $T^{0i}=T^{i0}$ and $T^{00}$ of the equations of motion. In general this will involve integrating a partial differential equation, but for the special cases that we have studied, this ``conserving completion'' of the equations of motion involves solving only an algebraic equation, or an ordinary  differential equation that can be integrated in closed form.

    The method of getting equations of motion just described contrasts with that used in effective theories of inflation \cite{effinfl} and effective theories of dark energy \cite{effdark}, which both start from a four-space general coordinate invariant action, and then introduce a unitary gauge coordinate fixing, which breaks time diffeomorphism symmetry, leaving only a three-space general coordinate invariance.  In these papers the energy-momentum tensor is constructed by varying the action with respect to the full $g_{\mu\nu}$, and reflecting the full invariance of the starting action, this tensor is  covariantly consered. The authors of both papers analogize their procedure to spontaneous symmetry breaking in gauge theories, where the vacuum breaks gauge symmetry, but the usual differential equations of conservation coming from the Noether theorem remain valid.  In this analogy, our action is like intrinsic, as opposed to spontanous, breaking of symmetry in gauge theory, and our model lies outside the general framework of \cite{effdark}, as we discussed in more detail in Appendix D of the  paper \cite{adler1d}.\footnote{Spontaneous symmetry breaking is associated with the appearance of a massless Nambu-Goldstone scalar mode, which has an analog in the analysis of \cite{effdark}; intrinsic symmetry breaking does not lead to such a massless scalar, and none appears in our analysis of the consequences of Eq. \eqref{eff}.}

\item{\bf Spherically Symmetric Solutions}  As a first application of Eq. \eqref{eff},  spherically symmetric vacuum solutions were studied in \cite{adler1c}. In this application the conserving completion involves only solving an algebraic equation, not a differential equation.  The results of solving the Einstein equations, including the action of Eq. \eqref{eff}, show that there is no black hole horizon where $g_{00}$ vanishes, but the change from the usual exterior Schwarzschild metric is confined to within a distance ${10}^{-17} ~~{\cal M}^2 ~ {\rm cm}$ of the nominal horizon, with  ${\cal M}$ the black hole mass in solar mass units.  So the usual astrophysics of black holes should not be affected, but there could be significant consequences, still to be studied, for the so-called ``black hole information paradox''.

\item{\bf Phenomenology, Cosmological Small Fluctuation Theory, and Absence of a Propagating Scalar Mode}

To set up a phenomenology to compare the cosmological predictions of the actions of Eq. \eqref{cosm} and \eqref{eff}, in \cite{adler1d}  we studied a dark energy action that is a parameterized linear combination of the two,
\begin{equation} \label{ansatz}
S_{\Lambda} = (1-f)  S_{\rm cosm}+ f  S_{\rm eff}
= -\frac{\Lambda}{8\pi G} \int d^4x (^{(4)}g)^{1/2} [1-f + f(g_{00})^{-2}]~~~~,
\end{equation}
so that $f=0$ corresponds to only a standard cosmological constant, and $f=1$ corresponds to only an apparent cosmological constant
arising from a frame dependent effective action. When $g_{00}=1+h_{00}\equiv 1+E$, with $E$ a perturbation treated to first order, Eq. \eqref{ansatz} reduces to
\begin{equation}\label{ansatz1}
S_{\Lambda}=S_{\rm cosm}+\frac{\Lambda}{4\pi G} \int d^4x (^{(4)}g)^{1/2} f E ~~~,
\end{equation}
allowing the study of nonzero $f$ to be treated by methods of standard small fluctuation theory \cite{weinberg}  around an unperturbed FLRW background calculated using Eq. \eqref{cosm} as the dark energy action.  This leads to additions, proportional to $f$, to the standard small fluctuation equations of motion.  Using the residual gauge invariance under spatial diffeomorphisms, which allows one to eliminate the fluctuation amplitude $B$, we  showed in \cite{adler1d} that even with $f\neq 0$, the modified  small fluctuation equations in $B=0$ gauge do {\it not} admit propagating scalar gravitational wave solutions.  This is an important  check on the $f\neq 0$ theory, since observational evidence is consistent with gravitational waves being purely tensor, with no  scalar component.

\item{\bf Numerical Solution of the Small Fluctuation Equation,  and Coordinate Time Formulation of Cosmography}

 Finally,  in \cite{adler1} [hereafter  (I)] we introduced a fluctuation amplitude $\Phi$ around the FLRW background metric,  in terms of which the perturbed line element takes the form
 \begin{align}\label{phipsi}
 ds^2=&[1+2\Phi(t)]dt^2 -[a^2(t)/\big(1-2\Phi(0)\big)][1-2\Phi(t)]d{\vec x}^{\,2}~~~\cr
 =&d\tau^2 -a^2(t)[1-2\big(\Phi(t)-\Phi(0)\big)]d{\vec x}^{\,2},~~~\cr
 \end{align}
with $a(t)$ the standard FLRW expansion factor,  and with the  proper time $\tau$ related to the coordinate time $t=x^0$ by
\begin{equation}\label{propertime}
d\tau=[1+\Phi(t)] dt=\big(1+\Phi(0)\big)[1+\Phi(t)-\Phi(0)]dt~~~.
\end{equation}
The rescaling by the constant factor $1+\Phi(0)$ in $d\tau$ is absorbed in the definition of the Hubble constant measured at early time, and the factor $1-2\Phi(0)$ rescaling $a(t)$ drops out of the time evolution equation for $\Phi$, which is independent of the scale of $a(t)$.\footnote{In writing Eq. \eqref{phipsi} we incorporated the leading order relation $\Phi=\Psi$, which is independent of $a(t)$ and thus its normalization when the anisotropic inertia $\pi^S$ vanishes.  Details of the elimination of the fluctuation amplitude $\Psi$ in terms of $\Phi$ are given in (I).}  The effect of these rescalings in the final formulas given below is that $\Phi(t)$ always appears in the subtracted form $\Phi(t)-\Phi(0)$, which  guarantees that the perturbation corrections do not appear in the early time formulas used in the cosmic microwave background analysis.
  In (I), after approximating a numerical solution of the dimensionless form of the $\Phi$ evolution equation, for the special case $f=1$, by a quadratic function of time, derivations of cosmographical formulas were carried out using the coordinate time $t$ ,  taking account of the above constant rescalings.
\end{itemize}

Having sketched out this background of earlier work, we can now state the purposes of the present paper.  They are (i) to redo the analysis of (I) entirely in terms of proper time, (ii) to give an additional formula for the squared effective Hubble parameter not given in (I),  (iii) to give analytical results for $\Phi$ which yield detailed information about its dependence on general $f$, and (iv) to express all results in terms of $\Phi$ without first making a quadratic approximation.   Expressed in terms of proper time, the
$f$-dependent evolution equation for $\Phi$ takes the following form.
Introducing the dimensionless time variable $x=\frac{3}{2} \surd{\Omega_{\Lambda}} H_0^{\rm Pl} \tau$, with $\tau$ the proper time,\footnote{We follow as closely as possible the notational conventions of (I), where as explained above $t$ was used for the coordinate time employed  and $\tau$ was used for the physical proper time. When the rescalings introduced in (I) are taken into account, the definition of $x$ there agrees with the one in this paper at early times.}
with $\Omega_{\Lambda}$ the cosmological fraction, and with $H_0^{\rm Pl}$ the Hubble constant as measured by Planck \cite{planck}, the evolution equation  for the metric perturbation $\Phi$ takes the form
\begin{equation}\label{evolution2}
\frac{d^2 \Phi}{dx^2}+\frac{8}{3} \coth(x) \frac{d \Phi}{dx} =\frac{4}{3}(2f-1) \Phi~~~.
\end{equation}
When $f=0$, this equation agrees with the one given in standard cosmology references
\cite{ma}, \cite{mukhanov}.
At the present era $x$ takes the value $x_0={\rm arcsinh}\big((\Omega_\Lambda/\Omega_m)^{1/2}\big)\simeq 1.169$  in the unperturbed FLRW cosmology, so to get $\Phi(x_0)$ we must integrate Eq. \eqref{evolution2} from $x=0$ to $x=x_0$, which was done numerically in (I) for $f=1$.

We see that the effect of changing the value of $f$ is to change the coefficient of the right-hand side of Eq. \eqref{evolution2}, with $f=1$ reversing the sign from the value at $f=0$.  Since the cosmological constant $\Lambda$ is very small on the distance scale of the early universe, the coefficient  of the right-hand side only significantly affects cosmology at late times when dark energy leads to acceleration of the expansion of the universe. Thus our analysis,  when applied to the much discussed  ``Hubble tension'' \cite{riess1}, \cite{hubble},  is a ``late time'' model.  In (I) we noted obstacles faced by ``late time''  models as discussed in \cite{lemos}, arising from BAO measurements of the Hubble constant \cite{alam1}.  We also referenced in (I) a review of the requirements that ``early time'' models \cite{knox} must obey, as well as other literature.  We believe that the currently available experimental data does not decisively distinguish between classes of models, and that it very likely will be several years before there are results that do so.  In anticipation of these, we present in this paper further formulas that will be useful for tests of our model when more extensive data is available.

This paper is organized as follows.  In Sec. II and Appendix A of this paper, we give analytic methods for obtaining  $\Phi(x)$ for general $f$. In Sec. III we rederive the principal results of Sec. V of (I) using formulas expressed directly in terms of proper time, thus avoiding reference to the auxiliary coordinate time and rescalings used in (I), and also avoiding using a quadratic approximation to $\Phi$.  Based on these formulas,  we give a compact formula for the squared Hubble parameter versus redshift, which should be useful in future comparisons with experiment.    In Sec. IV, assuming the model of Eq. \eqref{ansatz},  we discuss the implications of various Hubble tension outcomes for the structure of the dark energy action.  In particular, we use the results of Sec. II  to discuss the extent to which an observation of a late time increase of the Hubble parameter could be interpreted as evidence for frame dependence of the dark energy action. In Appendix A we show that an exact solution for $\Phi$ can be given in terms of the well-studied hypergeometric function.  In Appendix B we reexpress our results as a modified dark energy equation of state in the standard FLRW cosmology.

\section{Analytic formulas for $\Phi(x)$  }

It is convenient now to write $\Phi(x)=\Phi(0)\hat \Phi(x)$, so that $\hat \Phi(0)=1$, and $\Phi(x)-\Phi(0)=\Phi(0)[\hat \Phi(x)-1].$
In (I) we determined $\hat\Phi(x)$ by numerical integration for the special case when  the parameter $f$ introduced in Eq. \eqref{ansatz} is unity, corresponding to a Weyl scale invariant dark energy action.   Alternatively, we can get analytic approximations to $\hat \Phi(x)$, including the dependence on general $f$, by solving Eq. \eqref{evolution2} by power series expansion.  Writing
\begin{equation}\label{expan}
\hat\Phi(x)= 1+\hat C x^2 +\hat D x^4 + O(x^6)~~~,
\end{equation}
substituting into Eq. \eqref{evolution2}, and equating coefficients of like powers, we get the results
\begin{align}\label{coeffs}
\hat C =& \frac{2}{11} (2f-1)~~~,\cr
\hat D =& \frac{2}{187} (2f-1)(2f-7/3)~~~,\cr
\hat \Phi(x)=&1+\frac{2}{11} (2f-1)x^2\left[1+\frac{1}{17}(2f-7/3)x^2\right]+O(x^6)~~~.\cr
\end{align}
When $f=1$,  these give at $x_0 \simeq 1.169$,
\begin{align}\label{seriesval}
\hat C=& \frac{2}{11}=0.1818~~~,\cr
\hat D=&\frac{2}{187}\left(-\frac{1}{3}\right)=-0.00357~~~,\cr
\hat \Phi(x_0)-1\simeq & \hat C x_0^2 + \hat D x_0^4 =0.242~~~,\cr
\end{align}
very close to the value $\hat\Phi(x_0)-1=0.244$ found in (I) by numerical integration.  So the first two terms of the series expansion solution give an accurate approximation to $\hat \Phi(x)$, and have the virtue of giving the dependence on the parameter $f$.  When $f=0$, corresponding to the standard dark energy action, we have instead
\begin{align}\label{seriesval1}
\hat C=& -\frac{2}{11}=-0.1818~~~,\cr
\hat D=&\frac{2}{187}\left(\frac{7}{3}\right)=0.02496~~~,\cr
\hat \Phi(x_0)-1\simeq & \hat C x_0^2 + \hat D x_0^4 =-0.202~~~.\cr
\end{align}

 One can do even better than these power series expansions, by a change of variable that converts Eq. \eqref{evolution2} to a hypergeometric equation.  Details of this are given in Appendix A, and the result is the elegant formula
\begin{align}\label{hyper1}
\hat\Phi(x)=&(1-u)^b \,{}_2F_1(a,b,c; u)~~~,\cr
a=&\frac{1}{2}+b~~~,\cr
b=&\frac{1}{3}[2+(6f+1)^{1/2}]~~~,\cr
c=&\frac{11}{6}~~~,\cr
u=&{\rm tanh}^2(x)~~~.\cr
\end{align}
When this is expanded in powers of $x$, it agrees with the series coefficients  derived above.  In Appendix A we show that the formula of Eq. \eqref{hyper1} is an even function of the square root $(6f+1)^{1/2}$, which is why no square root appears in the series expansions, and  why the sign chosen for this square root in deriving the hypergeometric function formula is irrelevant.

\section{Formulas expressed in terms of proper time $\tau$ and the exact, unapproximated $\hat \Phi$}
In terms of the proper time $\tau$, the line element for our model takes the form
\begin{equation}\label{line}
ds^2=d\tau^2-\psi^2[\tau] d{\vec x}^{\,2}~~~,
\end{equation}
with $\psi[\tau]$ given by the following formulas, which are the extensions to an exact $\hat \Phi(x)$ of the quadratic approximation formulas in (I),
\begin{align}\label{psieq}
 \Delta(x)\equiv &\Phi(0)[\hat \Phi(x)-1] ~~~,\cr
\psi[\tau]=&a[\tau]\left[1-\Delta(x)\right]~~~,\cr
a[\tau]=&\left(\frac{\Omega_m}{\Omega_{\Lambda}}\right)^{1/3} \Big(\sinh(\hat x[\tau])\Big)^{2/3}~~~,\cr
\hat x[\tau]=&x-\int_0^x du \Delta(u)~~~,\cr
x=&\frac{3}{2} \surd{\Omega_\Lambda}H_0^{\rm Pl}\tau~~~.\cr
\end{align}
In terms of $a[\tau]$, the redshift $z_{\rm eff}$ in our model, for light emittted at proper time $\tau$ and observed at the present proper time $\tau_0$, is given by
\begin{equation}\label{red}
1+z_{\rm eff}=\frac{\psi[\tau_0]}{\psi[\tau]}=\frac{1}{\psi[\tau]} = \frac{1}{a[\tau][1-\Delta(x)]}~~~.
\end{equation}
Here $\Omega_\Lambda$ and $\Omega_m$ are the dark energy and matter fractions (which sum to unity in the matter dominated era), $H_0^{\rm Pl}\simeq 67.27 {\rm km}\rm {s}^{-1} {\rm Mpc}^{-1}$ is the Planck \cite{planck} Hubble constant, and $\Phi(0)$, which is treated as a small quantity to first order,  is the sole parameter of the model.  Thus our model extends the six parameter $\Lambda{\rm CDM}$ model to a seven parameter model, with $\Phi(0)$ as the seventh parameter.   When $\Phi(0)=0$, and also in the small $x$ limit, the above equations reduce to standard equations of the FLRW cosmology.  To reduce notational clutter, we have incorporated $\Phi(0)$ into the definition of $\Delta(x)$, and so $\Delta(x)$ is a small quantity treated to first order.  In the formulas given in (I), where we used the quadratic approximation $\hat \Phi(x)\simeq 1 + C(x/x_0)^2$, with $C=0.244$, we had $\Delta(x)  \simeq \Phi(0) C (x/x_0)^2$,
$\int_0^x du \Delta(u)=(x/3) \Delta(x)$, and $d\Delta(x)/dx=(2/x)\Delta(x)$.

Corresponding to the line element of Eq. \eqref{line}, the Hubble parameter $H_{\rm eff}[\tau]$ is given by
\begin{align}\label{hubeff}
H_{\rm eff}[\tau]=&\frac{d\psi[\tau]/d\tau}{\psi[\tau]}\cr
=&\frac{d\psi[\tau]/dx }{\psi[\tau]} ~dx/d\tau\cr
=&\left[\frac{da[\tau]/d\hat x[\tau]}{a[\tau]}~d\hat x[\tau]/dx-\frac{d\Delta(x)}{dx}\right]\frac{3}{2}\surd{\Omega_\Lambda}H_0^{\rm Pl}  \cr
=&H_0^{\rm Pl}\surd{\Omega_\Lambda} {\rm coth}(\hat x[\tau])\left[1-\Delta(x)-\frac{3}{2} {{\rm tanh}(x)}\frac{d\Delta(x)}{dx}
\right]~~~.\cr\end{align}
Dividing by $H_0^ {\rm Pl}$, squaring, and using $\coth^2(\hat x[\tau])=1+1/\sinh^2(\hat x[\tau])$, and then using Eqs. \eqref{psieq} and \eqref{red} to eliminate $\sinh^2(\hat x[\tau])$ in terms of the cube of  $(1+z_{\rm eff})[1-\Delta(x)]$, we get the result
\begin{align}\label{final}
\left(\frac{H_{\rm eff}[\tau]}{H_0^{\rm Pl}}\right)^2=& \tilde{\Omega}_m (1+z_{\rm eff})^3 +\tilde{\Omega}_{\Lambda} ~~~,\cr
\frac{\tilde{\Omega}_m}{\Omega_m}=& 1-5\Delta(x)-3\,{\rm tanh}(x)\frac{d\Delta(x)}{dx}~~~,\cr
\frac{\tilde{\Omega}_\Lambda}{\Omega_\Lambda}=&1-2\Delta(x)-3 \,{\rm tanh}(x)\frac{d\Delta(x)}{dx}~~~,\cr
\end{align}
with $\tilde{\Omega}_m$ and $\tilde{\Omega}_\Lambda$ redshift-dependent effective matter and dark energy densities.
To calclate $x$ in the above formulas from $z_{\rm eff}$, it suffices to use the zeroth order formulas
\begin{align}\label{xtau}
x = &{\rm arcsinh}(s)=\log(s + (s^2+1)^{1/2})~~~,\cr
s=&\left(\frac{\Omega_\Lambda}{\Omega_m}\right)^{1/2}\frac{1}{(1+z_{\rm eff})^{3/2}}~~~.\cr
\end{align}

Returning to Eq. \eqref{hubeff}, to calculate the present ratio $H_{\rm eff}({\rm present})/H_0^{\rm Pl}=H_{\rm eff}[\tau_0]/H_0^{\rm Pl}$,
we first need to calculate the present proper time $\tau_0$, which is determined by the condition $\psi[\tau_0]=1$. Equivalently, we have to calculate $x_{\tau_0}=(3/2) \surd{\Omega_\Lambda}H_0^{\rm Pl} \tau_0$, and to do this we proceed as follows.  We begin by writing
$\hat x[\tau_0]=x_0+ \Delta \hat x$,  so that the first order perturbation $\Delta \hat x$ is fixed by
\begin{align}\label{calc1}
[1+\Delta(x_0)]\psi[\tau_0]=&1+\Delta(x_0)=a[\tau_0]\cr
=&\left(\frac{\Omega_m}{\Omega_{\Lambda}}\right)^{1/3} \left(\sinh(\hat x[\tau_0])\right)^{2/3}\cr
=&\left(\frac{\Omega_m}{\Omega_{\Lambda}}\right)^{1/3} \left(\sinh(x_0+\Delta \hat x)\right)^{2/3}\cr
=&1+(2/3)\coth(x_0)\Delta \hat x~~~,\cr
\end{align}
which using $\coth(x_0)=1/\surd{\Omega_\Lambda}$ gives
\begin{equation}\label{calc2}
\Delta \hat x=(3/2) \surd{\Omega_\Lambda} \Delta(x_0)~~~.
\end{equation}
We then invert the relation between $x$ and $\hat x[\tau]$ to give
\begin{align}
x_{\tau_0}=&\hat x[\tau_0]+\int_0^{x_0} du \Delta(u)\cr
=&x_0+(3/2)\surd{\Omega_\Lambda}\Delta(x_0)+\int_0^{x_0} du \Delta(u)~~~.\cr
\end{align}
Multiplying through by $2/(3\surd{\Omega_\Lambda}H_0^{\rm Pl})$, this is equivalent to
\begin{equation}\label{calc3}
\tau_0=\tau_0^{\rm Pl}+
\frac{1}{H_0^{\rm Pl}}\left[\Delta(x_0)+\frac{2}{3\surd{\Omega_\Lambda}}\int_0^{x_0} du \Delta(u)\right]
\end{equation}
with $\tau_0^{\rm Pl}=2x_0/(3\surd{\Omega_\Lambda}H_0^{\rm Pl})=13.83 {\rm Gyr}$.

Returning now to the calculation of $H_{\rm eff}({\rm present})/H_0^{\rm Pl}$, from Eq. \eqref{hubeff} we get
\begin{equation}\label{hubeff1}
H_{\rm eff}({\rm present})/H_0^{\rm Pl}=
\surd{\Omega_\Lambda} {\rm coth}(x_0+\Delta \hat x)\left[1-\Delta(x_0)
-(3/2)\surd{\Omega_\Lambda}d\Delta(x_0)/dx_0\right]~~~.
\end{equation}
From the expansion
\begin{equation}\label{expans}
{\rm coth}(x_0+\Delta \hat x)=\frac{1}{\surd{\Omega_\Lambda} }[1+(3/2)(\Omega_{\Lambda}-1)\Delta(x_0)]
= \frac{1}{\surd{\Omega_\Lambda} }[1-(3/2)\Omega_m \Delta(x_0)]~~~,
\end{equation}
we get
\begin{equation}\label{hubeff2}
\frac{H_{\rm eff}({\rm present})}{H_0^{\rm Pl}}=1-\left[\left(\frac{3}{2}\Omega_m
+1\right)\Delta(x_0)+\frac{3}{2}\surd{\Omega_\Lambda}\frac{d\Delta(x_0)}{dx_0}\right]~~~.
\end{equation}
Putting in numerical values $\Omega_m=0.321$, $\Omega_{\Lambda}=0.679$ and  $x_0\simeq  1.169$, together with the $f=1$ values from (I),\,   $\hat\Phi(x_0)-1=0.244$ \, and \, $d\hat\Phi(x_0)/dx_0=0.409$, we get
\begin{equation}\label{hubeff3}
\frac{H_{\rm eff}({\rm present})}{H_0^{\rm Pl}}\simeq 1-0.867 \,\Phi(0)~~~,
\end{equation}
and so to fit a Hubble tension  of $\frac{H_{\rm eff}({\rm present})}{H_0^{\rm Pl}}\simeq 1.1$ with the scale invariant action of Eq. \eqref{eff} we would need to choose $\Phi(0) \simeq -0.115$.
For this value of $\Phi(0)$, the correction factors $\tilde\Omega_m/\Omega_m$ and $\tilde \Omega_{\Lambda}/\Omega_{\Lambda}$ in Eq. \eqref{final} are greater than unity by roughly 20 to 30 percent.  Thus it will be important to include these in analyses assessing the viability of our model.\footnote{An interesting feature of the latest eBOSS data release \cite{alam2} is that in the plots of Fig. 5 of \cite{alam2}, the contour in the $H_0$--$\Omega_m$ plane for $z<1$ data is substantially offset towards larger values of $\Omega_m$ and $H_0$ from the contour for $z>1$ data.  This raises the question of whether new physics involving a change in the effective $\Omega_m$ at small redshifts, such as predicted in our model,  could be playing a role.
I wish to thank Eva-Maria Mueller for helpful email correspondence about the eBOSS data; the interpretation suggested in this footnote is ours. }

\section{Implications of various Hubble tension outcomes for the structure of the dark energy action}

Since all formulas derived above depend on the unknown initial value $\Phi(0)$, our model does not predict a value for the Hubble tension.  However, going in the reverse direction from observed values of the Hubble parameter $H_{\rm eff}(\tau)$, we can make the following statements.
\begin{itemize}
\item  If the Hubble tension should turn out to be an artifact arising from mis-calibration of the distance ladder, as suggested in \cite{feeney}, or if the Hubble tension is real but arises exclusively from early time physics, as advocated in \cite{knox}, then our analysis is consistent with  $\Phi(0)=0$.  In this case the dark energy action of Eq. \eqref{eff} would be indistinguishable from the standard action of Eq. \eqref{cosm}, via cosmological measurements, through first order in perturbations.
\item  If the reported Hubble tension survives further testing, and has a late time component, then $\Delta(x_0)=\Phi(0) [\hat \Phi(x_0)-1]$
     must be negative. If data does not permit extraction of the quartic term in Eq. \eqref{expan}, then we approximate $\Delta(x_0)$ by the quadratic term
$\Phi(0)\hat C x_0^2 = \Phi(0)\frac{2}{11}(2f-1) x_0^2$ in giving allowed ranges of $f$.  If $\Phi(0)$ is negative, which could be argued to be plausible since it is an analog of the Newtonian gravitational potential \cite{wein}, then $\hat C$ has to be positive.  This requires
    $f>1/2$ and suggests presence of a frame dependent component in the dark energy action. On the other hand, if $\Phi(0)$ is positive, then $\hat C$ has to be negative, which requires $f<1/2$.
     This  range includes $f=0$ and so allows the dark energy action to be purely of the conventional form. Thus it would be of interest to try to find convincing arguments giving the expected sign of a nonzero $\Phi(0)$.
\item If future high accuracy data fits the expansion of Eq. \eqref{expan} and permits extraction of the coefficient of the quartic term, then from  the ratio
         \begin{equation}\label{extract}
         \frac{\hat D}{\hat C}=\frac{1}{17}(2f-7/3)
         \end{equation}
         one can determine $f$, independent of the value of $\Phi(0)$.
\item  Although we have focused on the predictions of our dark energy model for the expansion rate $H_{\rm eff}$, it will have cosmological consequences elsewhere which should be tested.  This will be an important focus for future work if fits to the Hubble tension data suggest that $f>0$ and $\Phi(0)\neq 0$. Other potential  applications of the cosmographical formulas given in this paper  include structure formation
    and the so-called $\sigma_8$ tension in clustering data \cite{sigma}, and comparisons of the ages of the oldest stars with the age of the universe \cite{age}.  The model we have proposed will be acceptable only if it fits all sets of cosmological data.

\end{itemize}

\section{Acknowledgement} We wish to thank the reviewer for pointing out that the differential equation for $\hat \Phi(x)$ is exactly solved by a hypergeometric function, and for many helpful suggestions on presentation.

\appendix

\section{Solution of the $\hat \Phi$ equation in terms of the hypergeometric function}

Dividing $\Phi(x)$ by the initial value $\Phi(0)$, Eq. \eqref{evolution2} reads
\begin{equation}\label{evolution3}
\frac{d^2 \hat\Phi}{dx^2}+\frac{8}{3} \coth(x) \frac{d \hat\Phi}{dx} -\frac{4}{3}(2f-1)\hat\Phi=0 ~~~,
\end{equation}
with the initial condition $\hat \Phi(0)=1$.  Making the change of variable
$u={\rm tanh}^2(x)$, this equation takes the form
\begin{equation}\label{evolution4}
4u(1-u)^2\frac{d^2\hat\Phi}{du^2} +2(1-u)(\frac{11}{3}-3u)\frac{d\hat\Phi}{du}-\frac{4}{3}(2f-1)\hat\Phi=0~~~.
\end{equation}
Now define a function $T(u)$ in terms of the hypergeometric function $w(u)\equiv{}_2F_1(a,b,c; u)$ by
\begin{equation}\label{Tdef}
w(u)=(1-u)^g T(u)~~~.
\end{equation}
Substituting this into the hypergeometric equation
\begin{equation}\label{hypereq}
u(1-u) \frac{d^2w}{du^2}+ [c-(a+b+1)u]\frac{dw}{du}-abw=0~~~,
\end{equation}
and multiplying by an overall factor $4(1-u)^{1-g}$, we find the following differential equation obeyed by $T(u)$,
\begin{equation}\label{Qeq}
4u(1-u)^2\frac{d^2T}{du^2}+4(1-u)[c-(a+b+1)u-2gu]\frac{dT}{du}
+4[g(g-1)u-ab\big(1-u-g(c-(a+b+1)u\big)]T=0~~~.
\end{equation}
Comparing this equation with Eq. \eqref{evolution4}, we find they agree in structure with the following choices of parameters $a,b,c,g$,
\begin{align}\label{param}
a=&\frac{1}{2}+b~~~,\cr
b=&\frac{1}{3}[2\pm(6f+1)^{1/2}]~~~,\cr
c=&\frac{11}{6}~~~,\cr
g=&-b~~~,\cr
\end{align}
giving the result for $\hat\Phi(x)$ quoted in Eq. \eqref{hyper1} of the text.\footnote{This solution can be obtained from Mathematica by the command: DSolve[\{u$^{\prime\prime}$[x] + (8/3) Coth[x] u$^{\prime}$[x] - (4/3) (2f - 1) u[x] == 0,
  u[0] == 1\}, u, x]~~.}  In that equation we chose the $+$ sign in $b$, but in fact this choice of sign is irrelevant.  This can be seen by using the so-called $Q$-form of the hypergeometric equation \cite{wikihyper} to write
\begin{equation}\label{Qform}
\hat\Phi(u)=u^{-11/12}(1-u)^{1/6}P(u)~~~,
\end{equation}
where $P(u)$ obeys the differential equation
\begin{align}\label{Peq}
&\frac{d^2P}{du^2}+Q(u)P(u)=0~~~,\cr
&Q(u)=\frac{u^2[1-(a-b)^2]+u[2c(a+b-1)-4ab] +c(2-c)}{4u^2(1-u)^2}~~~.
\end{align}
Writing $b=2/3+K$, with $K$ the square root term, we have
\begin{align}\label{Peqnum}
[1-(a-b)^2]=&3/4~~~,\cr
2c(a+b-1)-4ab=&-1/18-4K^2~~~,\cr
\end{align}
and so $P(u)$, and hence $\hat\Phi(u)$, is an even function of $K$.
 \footnote{A consequence of this is that the large $x$ form of $ \Phi(x)$ given in Eq. (23) of \cite{adler1} is an equal mixture of the two terms given there, i.e. $\Phi(x)= {\rm constant} \times (e^{\mu_+ x}  + e^{\mu_-x})$, with $\mu_{\pm}=-(2/3)[2\pm (6f+1)^{1/2}]$.}

\section{$H_{\rm eff}$ rewritten as a modified dark energy equation of state in the standard FLRW cosmology}

In the standard FLRW cosmology, for any component of the universe with pressure $p$, density $\rho$ and equation of state $p=w(z)\rho$, integrating the energy-momentum conservation equation
$\dot \rho =-3 (\dot a/a)(p+\rho)$ and using the redshift definition $1+z=1/a$ gives the result
\begin{equation}\label{densev}
\rho=\rho_0 e^{3 \int_0^z du [1+w(u)]/(1+u)}~~~.
\end{equation}
So in the matter-dominated era, with pressure-free matter  and dark energy of state $w_{\Lambda}(z)$, we get the standard FLRW cosmology formula \cite{tripathi}
\begin{equation}\label{densev1}
H^2(z)=H^2(0)\left[\Omega_m (1+z)^3+ \Omega_{\Lambda}\exp \left\{3 \int_0^z \frac{du}{1+u}[1+w_{\Lambda}(u)]\right\}\right]~~~.
\end{equation}
Comparing this equation with Eq. \eqref{final} \big(identifying $H_{\rm eff}$ with $H(z)$ and $z_{\rm eff}$ with $z$\big), we see that our model can be reinterpreted as the following dark energy equation of state in  Eq. \eqref{densev1},
\begin{align}\label{stateq}
U(z)\equiv&\exp \left\{3 \int_0^z \frac{du}{1+u}[1+w_{\Lambda}(u)]\right\}  ~~~,\cr
w_{\Lambda}(z)=&-1+\frac{1}{3}(1+z)\frac{d}{dz}\log U(z)~~~,\cr
U(z)=&\frac{1}{\Omega_{\Lambda}}\left[\frac{(\tilde\Omega_m(1+z)^3+\tilde\Omega_{\Lambda} )|_z}{(\tilde\Omega_m(1+z)^3+\tilde\Omega_{\Lambda})|_0}-\Omega_m (1+z)^3\right]~~~.\cr
\end{align}
As a consistency check, we note that  $U(0)=1$, and $U(z)=1$ when $\tilde\Omega_{\Lambda}=\Omega_{\Lambda}$ and $\tilde\Omega_m=\Omega_m$.

\end{document}